\documentclass[conference]{IEEEtran}
\IEEEoverridecommandlockouts

\usepackage{cite}
\usepackage{amsmath,amssymb,amsfonts}
\usepackage{algorithmic}
\usepackage{graphicx}
\usepackage{textcomp}
\usepackage{xcolor}
\usepackage{booktabs}
\usepackage{array}
\usepackage{tabularx}
\usepackage{multirow}
\usepackage{url}
\usepackage{hyperref}
\hypersetup{colorlinks=false, pdfborder={0 0 0}}

\def\BibTeX{{\rm B\kern-.05em{\sc i\kern-.025em b}\kern-.08em
    T\kern-.1667em\lower.7ex\hbox{E}\kern-.125emX}}

\begin{document}

\title{Functional Safety Analysis for 
Infrastructure-Enabled Depot Autonomy System}

\author{
Gaurav Pandey$^{1}$, Gregory Stevens$^{2}$, Henry Liu$^{2}$
\thanks{$^{1}$Gaurav Pandey is with The Department of Engineering Technology and Industrial Distribution Texas A\&M University, College Station, TX 77843, USA
        {\tt\small gpandey@tamu.edu}}
\thanks{$^{2}$Greg Stevens and Henry Liu are with MCity, University of Michigan Ann Arbor}
}

\maketitle

\begin{abstract}
This paper presents the functional safety
analysis for an Infrastructure-Enabled Depot Autonomy (IX-DA) system. The IX-DA system automates
the marshalling of delivery vehicles within a controlled depot
environment, navigating connected autonomous vehicles~(CAVs) between
drop-off zones, service stations (washing, calibration, charging,
loading), and pick-up zones without human intervention. We describe the
system architecture comprising three principal subsystems---the
connected autonomous vehicle, the infrastructure sensing and compute
layer, and the human operator interface---and derive their functional
requirements. Using ISO~26262-compliant Hazard Analysis and Risk
Assessment~(HARA) methodology, we identify eight hazardous events,
evaluate them across different operating scenarios, and assign Automotive
Safety Integrity Levels~(ASILs) ranging from Quality~Management~(QM)
to~C. Six safety goals are derived and allocated to vehicle and
infrastructure subsystems. The analysis demonstrates that high-speed
uncontrolled operation imposes the most demanding safety requirements
(ASIL~C), while controlled low-speed operation reduces most goals
to~QM, offering a practical pathway for phased deployment.
\end{abstract}


\section{Introduction}

\begin{figure}
    \centering
    \includegraphics[width=0.85\linewidth]{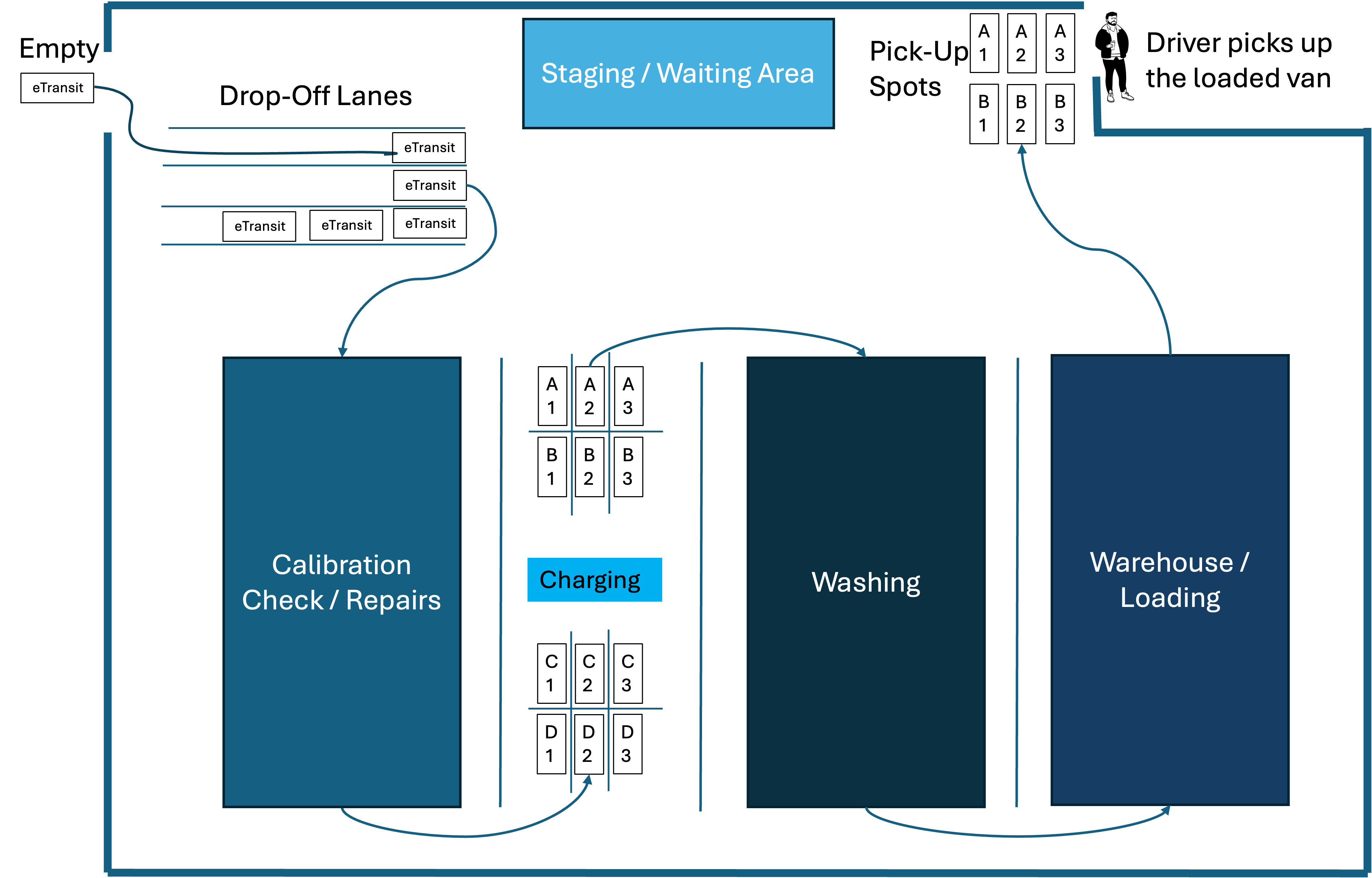}
    \caption{An IX-DA supported depot layout where the movement of vehicles between drop-off lanes,
washing bays, calibration stations, charging points, and loading
docks can be automated using an IX-DA system. The driver can drop off the empty vehicle in the drop-off zone and pick up a loaded vehicle from the pickup zone }
    \label{fig:depot-layout}
\end{figure}

Modern parcel and freight depots are characterised by repetitive,
low-complexity vehicle manoeuvres within a well-defined perimeter.
Delivery drivers must reposition vehicles between drop-off lanes,
washing bays, calibration stations, charging points, and loading
docks---a time-consuming process that accounts for a significant
fraction of non-productive dwell time. Automating these movements
through an IX-DA system can reduce dwell time, improve asset
utilisation, and remove drivers from hazardous yard environments (Fig~\ref{fig:depot-layout}).

Unlike on-road autonomy, depot marshalling takes place in a controlled,
GPS-denied environment with a predictable mix of participants (CAVs,
human-driven service vehicles, pedestrian workers) and bounded speed
regimes ($\leq 15$~mph). This bounded Operational Design Domain~(ODD)
makes infrastructure-centric autonomy---where a central compute layer
issues trajectory commands to each vehicle---both technically feasible
and cost-effective.

This paper documents the functional safety
analysis for an IX-DA prototype.
Section~\ref{sec:related} surveys related work.
Section~\ref{sec:arch} describes the system architecture and presents the functional requirements.
Section~\ref{sec:hara} details the hazard analysis and risk assessment (HARA).
Section~\ref{sec:sg} derives the safety goals and their ASIL ratings.
Section~\ref{sec:conc} concludes with deployment recommendations.

\section{Related Work}
\label{sec:related}



Automated Vehicle Marshalling~(AVM) has emerged as a recognised SAE
Level~4 use case for geofenced, low-speed environments. Schiegg
\emph{et al.}~\cite{schiegg2025} provide the most comprehensive survey
of AVM standardisation and deployment to date, covering the full
communication stack (V2X message formats), core motion-control mechanisms, and functional
safety principles. Their work situates depot logistics as one of three
primary AVM categories, alongside factory marshalling and automated
valet parking, and highlights ISO~23374-1~\cite{iso23374} and
SAE~J3292~\cite{saej3292} as the principal governing standards for
infrastructure-guided, driverless operation in prescribed areas.

Commercially, Embotech's AVM system, deployed across four BMW
manufacturing plants in Europe, demonstrates the viability of
infrastructure-centric autonomy. LiDAR sensors mounted in the factory
ceiling localise each vehicle to centimetre accuracy, and a central
autonomy stack transmits 150-byte trajectory packets at 10\,Hz over a
secure wireless link~\cite{embotech2025}. Unikie's Marshalling Solution
(UMS), validated in real operations within Volkswagen Group
Logistics at Emden, Germany, further confirms that infrastructure-based
guidance can handle the complex manoeuvres (narrow lanes and tight
parking bays), typical of industrial yards without
requiring onboard self-driving hardware~\cite{unikie2025}. Ford patented a method \cite{ford_marshalling_infra} in
which an infrastructure sensor network processes combined
roadside and on-board sensor signals to generate vehicle
commands. 



The broader literature on cooperative Vehicle--Infrastructure~(V2X)
systems motivates the IX-DA architecture at a theoretical level.
Extensive prior work~\cite{cvis2024, v2xsurvey2022, kimv2i2023, zou2022realtime, ying2024, zhang2024robust} demonstrates that
offloading perception and planning to roadside infrastructure can
compensate for the limited sensor range and compute budget of individual
vehicles, particularly in GPS-denied or occluded environments.
US~NHTSA estimates that
V2X adoption could prevent up to 615,000 crashes annually on public
roads~\cite{v2xsurvey2022}. In the depot context---where all vehicles
are instrumented, the network is private, and the ODD is tightly
bounded---these benefits are amplified. In a depot, every active vehicle is a known
agent, eliminating the partial-observability challenges that complicate
on-road V2X deployments. 

This paper makes two contributions:
(1)~it provides the first publicly documented
HARA for a delivery-depot IX-DA system, bridging the gap between
factory-floor AVM deployments and last-mile logistics;
(2)~it allocates safety goals between vehicle and
infrastructure subsystems in a manner consistent with emerging AVM
standards (ISO~23374-1, SAE~J3292), positioning it as
a reference implementation for future standardisation work.


\begin{figure}
    \centering
    \includegraphics[width=0.78\linewidth]{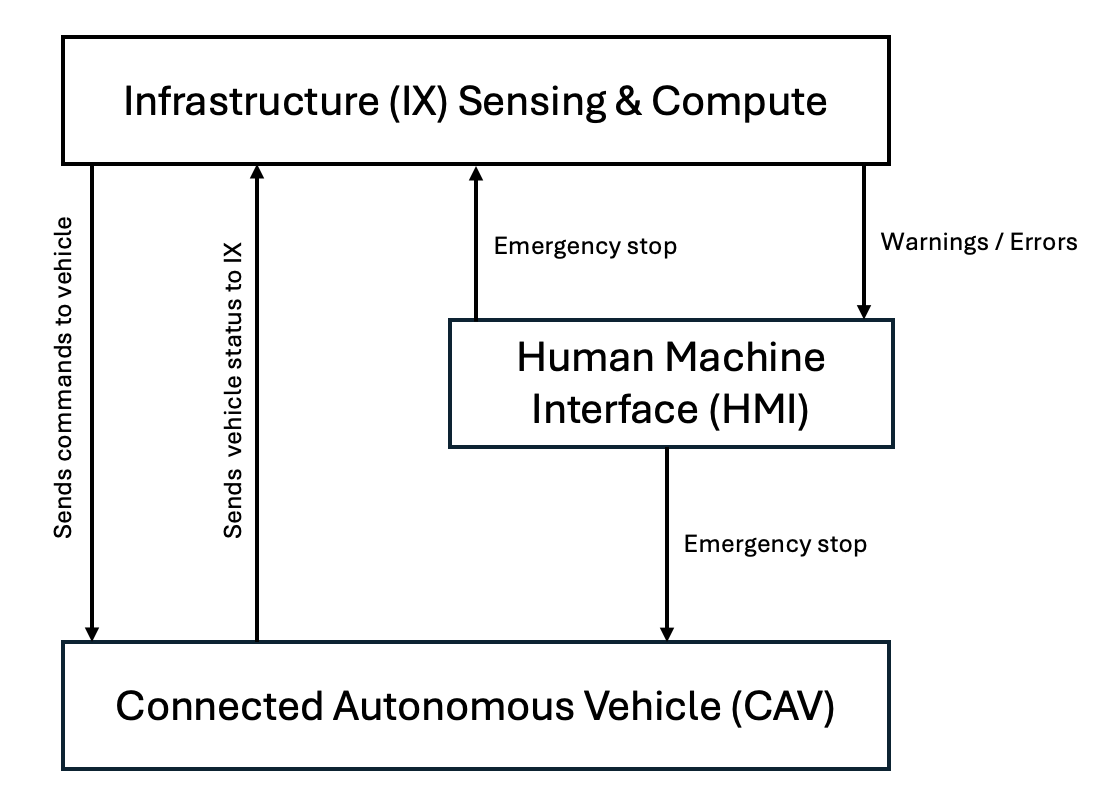}
    \caption{A high-level system boundary diagram and the typical communication between the three sub-systems (CAV, IX, HMI).}
    \label{fig-boundary-diagram}
\end{figure}

\section{System Architecture \& Functional Requirements}
\label{sec:arch}

The IX-DA system is partitioned into three principal subsystems: the
Connected Autonomous Vehicle~(CAV), the Infrastructure Sensing and
Compute layer~(IX), and the Human--Machine Interface~(HMI). A high-level system boundary diagram and the typical communication between the three sub-systems is shown in Fig~\ref{fig-boundary-diagram}. The CAVs are assumed to be equipped with capabilities for automatic charging and predictive maintenance, ensuring they operate at peak performance with minimal downtime. The infrastructure (IX) oversees the entire operation, employing advanced sensors and monitoring systems to track vehicle movements, environmental conditions, and security. It communicates with CAVs, sending commands and receiving status updates, and issues alerts to human operators regarding system status or emergencies. Users  have the ability to oversee operations in real-time, respond to alerts, initiate emergency protocols, and perform manual interventions or maintenance as necessary. Based on the relationship between these characteristics, more detailed functional requirements are discussed below.

\subsection{Connected Autonomous Vehicle (CAV)}
\label{cav-req}
In this work we assume that each CAV is a modified delivery van retaining its factory-fit brake,
steer, and throttle actuators, augmented with: (a)~an on-board sensor
suite (radar, camera) for short-range obstacle detection;
(b)~a V2I communication module (IEEE~802.11p / C-V2X); and (c)~a
Vehicle Control Unit~(VCU) that executes IX-provided trajectories at
10\,Hz. The detailed functional requirement for the CAV sub-system are listed below:
\subsubsection{Automatic Obstacle Detection and Collision Avoidance (AODCA)}
The vehicle must have an AODCA system that detects and tracks other vehicles, pedestrians, and unidentified obstacles that are within the stopping distance of the host vehicle, and sends a signal to VCU.

\subsubsection{Vehicle Control Unit}
The vehicle control unit shall execute the following planned actions:
\begin{itemize}
    \item Follow the path provided by the infrastructure.
    \item Activate automatic emergency braking (AEB) if a signal is received from AODCA.
    \item Stop the vehicle if the communication with the infrastructure is broken or delayed by more than 3 seconds. This time is calculated assuming the max speed to be 15mph and on-board sensing range for the vehicle to be 10m. The vehicle will travel 10m in 3 seconds at the speed of 15mph.
    \item	Stop the vehicle if there is significant failure in the vehicle system (e.g. brake system loss, power loss etc).
\end{itemize}
\subsubsection{Vehicle Communication System}
The vehicle communication system shall enable effective interaction with infrastructure and other connected entities.
\begin{itemize}
    \item The vehicle communication system shall receive trajectory to be followed at the rate of 10 fps, emergency stop command, commands to enter/exit a station from infrastructure.
    
    \item The vehicle communication system shall send the current state of vehicle (position, speed, drive mode, warnings, light status, door status, sensing data) to the infrastructure:

    \item The vehicle communication system shall implement robust cybersecurity measures to protect against unauthorized access and ensure the integrity of communication.
\end{itemize}


\subsection{Infrastructure Sensing and Compute (IX)}

We assume a network of sensors (LiDAR, cameras) that provide a complete coverage of the depot floor connected to a centralised
compute server. The detailed requirements for IX are listed below:
\subsubsection{Sensor}
The infrastructure system shall be equipped with advanced sensing technologies to detect and track dynamic and static elements within its environment.
\begin{itemize}
    \item The infrastructure sensing system shall detect all vehicles and pedestrians (estimate the speed, direction, and behavior of the detected vehicles and pedestrians) within the environment all the time.

    \item	The infrastructure sensing system shall detect any unidentified objects within the environment.

    \item The infrastructure sensing shall detect any operational hazards (like fire, smoke, earthquake) that prevent safe operation of the depot.

\end{itemize}

\subsubsection{Prediction}
The infrastructure system shall utilize predictive models to forecast potential changes and risks within the environment.
The infrastructure system shall have knowledge of future positions of all the vehicles and pedestrians in the environment.

\subsubsection{Planning}
The infrastructure system shall manage traffic and resources within the environment.
\begin{itemize}
    \item 	The infrastructure system shall plan a safe maneuver for all the vehicles within the environment at any given time.
\item	The infrastructure system shall manage the traffic flow within the environment such that there is no congestion at any given time.
\item	The infrastructure system shall manage the deployment of infrastructure resources, such as traffic signals and information boards, to support vehicle and pedestrian safety and efficiency.
\item	The infrastructure system shall implement comprehensive protocols for managing emergency situations (fire, flood, earthquake, accidents) to safeguard both human and vehicular traffic
\end{itemize}

\subsubsection{Communication}
The infrastructure system shall onboard any new vehicle prior to operation and maintain the state of the system for the last 10 seconds of operation. 

\begin{itemize}
    \item 	The infrastructure system shall validate that any new vehicle registered with the IX-DA system meets all the functional requirements described in section \ref{cav-req} above.
	
    \item The infrastructure system shall maintain a rolling buffer of the entire state of the depot for the last 10 sec so that it can resume operations from the last state in case of any power failures to the infrastructure.

\end{itemize}


\subsection{Human--Machine Interface (HMI)}

Three HMI touchpoints are assumed: (i)~a server dashboard for the
depot operator (ii)~a mobile application for drivers; and (iii)~physical emergency-stop buttons. 
\subsubsection{Server dashboard}
The system shall have a user interface for the human operator monitoring the entire depot. The server dashboard shall display the state (working/error) of all the sensors, the current position of all the vehicles within the depot, the errors of the vehicle within the depot, and the current planned path for all the vehicles. It shall also provide visual/audio notification in case any vehicle deviates from the planned path and allow the human operator to issue emergency stop command to any vehicle.

\subsubsection{Mobile application for Drivers}
The system shall have a user interface in the form of a mobile application for the drivers dropping and picking up the vehicle from the depot. The driver app shall authenticate the driver before checking in a vehicle in the drop-off zone. It shall also notify the driver about when and which vehicle she can check-out from the pick-up location.
It shall also authenticate the driver before allowing the driver to check-out a loaded vehicle.

\subsubsection{Physical Emergency Stop}
The system shall have easily accessible physical buttons, in locations of pedestrian traffic, that can be pressed by human workers in the depot to stop all the vehicles within a 10m radius.




\section{Hazard Analysis and Risk Assessment (HARA)}
\label{sec:hara}

We adopted the  hazard analysis and risk assessment procedure for an automated unmanned vehicle demonstrated in Stolte \emph{et al.} \cite{torben2017}. Each functional requirement is examined using four guide words---\emph{Loss
of Function}, \emph{More Than Intended}, \emph{Less Than Intended}, and
\emph{Intermittent}---to derive hazardous events. Table~\ref{tab:hazards}
lists the eight identified hazards.

\begin{table}[!t]
  \caption{Identified Hazards}
  \label{tab:hazards}
  \centering
  \renewcommand{\arraystretch}{1.15}
  \begin{tabular}{>{\bfseries}c p{3.9cm} p{3.cm}}
    \toprule
    ID & Cause & Hazardous Event \\
    \midrule
    H1 & Loss of vehicle AODCA
       & Collision with pedestrian \\
    H2 & Loss of vehicle braking
       & Collision with pedestrian \\
    H3 & Unintended acceleration
       & Collision with pedestrian \\
    H4 & Loss of V2I communication. No trajectory update
       & Collision with pedestrian \\
    H5 & Intermittent V2I communication. Delayed trajectory update
       & Collision with pedestrian \\
    H6 & Loss of IX sensing. Infrastructure blind to obstacles
       &  Collision with pedestrian\\
    H7 & Faulty IX prediction. Incorrect trajectory update
       & Collision with pedestrian \\
    H8 & Emergency stop unavailable. No intervention on critical fault
       &  Collision with pedestrian\\
    \bottomrule
  \end{tabular}
\end{table}

Risk is assessed for the identified hazardous events across two parameter dimensions.


\textbf{Operating Modes:} \emph{Nominal Speed (NS)} ($<10$~mph, stopping
distance $<10$~m, low collision impact); \emph{High Speed (HS)}
(10--25~mph, stopping distance $>10$~m, potentially life-threatening
impact at 25~mph).

\textbf{Traffic Situations:} \emph{Controlled (C)}---human access to
vehicle operating zones is restricted; \emph{Uncontrolled (UC)}---no
restrictions on pedestrian movement within the depot.

Severity~(S), Exposure~(E), and Controllability~(C) classes are
assigned per ISO~26262-3, and ASIL is calculated from the standard's
determination matrix. 


\section{Safety Goals}
\label{sec:sg}

Six safety goals (SG1--SG6) are derived from the worst-case ASIL
across all hazardous events and operating scenarios.
Table~\ref{tab:sg} presents each goal with its ASIL rating across the
three canonical scenarios: Nominal~Speed / Controlled~(NS/C), High~Speed
/ Controlled~(HS/C), and High~Speed / Uncontrolled~(HS/UC).

\begin{table*}[!t]
  \caption{Safety Goals and ASIL Ratings}
  \label{tab:sg}
  \centering
  \renewcommand{\arraystretch}{1.2}
  \begin{tabular}{>{\bfseries}c p{9.8cm} c c c c}
    \toprule
    Goal & Description & NS/C & HS/C & HS/UC & Assigned To \\
    \midrule
    SG1 & Ensure the vehicle has an independent AODCA system that
          detects obstacles within stopping distance and avoids
          collisions.
        & QM & QM & B  & Vehicle \\
    SG2 & Ensure the vehicle braking system achieves immediate full
          stop and is robust to single-point failures.
        & QM & QM & B  & Vehicle \\
    SG3 & Ensure maximum allowed acceleration and speed are limited to
          pre-defined values.
        & A  & A  & C  & Vehicle \\
    SG4 & Ensure V2I communication is robust and continuously monitored
          for disconnections.
        & QM & QM & A  & IX / Vehicle \\
    SG5 & Ensure the IX sensing system detects static and dynamic
          obstacles in the vehicle's path to avoid or mitigate
          collisions.
        & QM & A  & C  & IX \\
    SG6 & Ensure the IX system has an emergency-stop mechanism available
          at all times.
        & QM & A  & C  & IX \\
    \bottomrule
  \end{tabular}
\end{table*}

\subsection{Key Observations}

The most demanding requirements are SG3 (speed limiting), SG5 (IX sensing), and SG6 (emergency stop) at ASIL~C in
the HS/UC scenario, driven by S3 severity (life-threatening injury at
25~mph) combined with high exposure and low controllability. SG1 and
SG2 (on-board AODCA and braking) reach only ASIL~B in HS/UC because
the on-board AEB provides a degree of controllability even without
infrastructure intervention. SG4 (V2I robustness) reaches ASIL~A only
in HS/UC, reflecting the increased reliance on infrastructure at higher
speeds. For all six goals the NS/C scenario yields QM or ASIL~A at most,
providing a low-cost entry point. This motivates a phased deployment
strategy: begin with controlled, low-speed operation (NS/C),
progressively extend to HS/C once controlled-environment infrastructure
is validated, and only then address the more demanding HS/UC regime.

\subsection{Allocation}

SG1, SG2, and SG3 are allocated to the vehicle because they can be
addressed by on-board hardware and software independent of
infrastructure. SG4 is jointly allocated to the IX and the vehicle.
SG5 and SG6 are allocated to the IX because they require system-wide
awareness not available to individual vehicles.

\section{Conclusion}
\label{sec:conc}

This paper presents a structured functional
safety analysis for the IX-DA system. The architecture
decomposes into three well-defined subsystems---CAV, IX, and HMI. Eight hazardous events were
identified; HARA per ISO~26262-3 yielded
six safety goals with ASILs ranging from QM to~C depending on operating
scenario. The principal finding is that controlled low-speed operation is
achievable with modest safety integrity requirements (QM/ASIL~A),
enabling early commercial deployment. High-speed or uncontrolled
operation demands ASIL~B--C measures, particularly speed limiting~(SG3)
and IX sensing~(SG5), and should be deferred until appropriate hardware
and software validation is completed. Future work includes the development of the Safety Concept (allocating
safety requirements to architectural elements), failure mode effect and analysis (FMEA) at component level,
and hardware-in-the-loop validation of all the intended functions.



\bibliographystyle{IEEEtran}
\bibliography{refs}

\end{document}